\documentclass[aps,prl,twocolumn,nofootinbib]{revtex4-1}
\usepackage{graphicx}
\usepackage{bm}
\usepackage[colorlinks=true]{hyperref}
\usepackage{amsmath}
\usepackage{amsfonts}
\DeclareMathOperator{\hodge}{*\!}

\begin{document}
\title{Quantum statistics of vortices from a dual theory of the XY ferromagnet}

\author{Sayak Dasgupta}
\thanks{S. D. and S. Z. contributed equally to this work.}
\author{Shu Zhang}
\thanks{S. D. and S. Z. contributed equally to this work.}
\author{Ibrahima Bah}
\author{Oleg Tchernyshyov}
\affiliation{Department of Physics and Astronomy, Johns Hopkins University, Baltimore, Maryland, USA}

\begin{abstract}
We extend the well-known mapping between the easy-plane ferromagnet and electrostatics in $d=2$ spatial dimensions to dynamical and quantum phenomena in a $d=2+1$ spacetime. Ferromagnetic vortices behave like quantum particles with an electric charge equal to the vortex number and a magnetic flux equal to the transverse spin of the vortex core. Vortices with half-integer core spin exhibit fermionic statistics. 
\end{abstract}

\maketitle

Dualities are versatile tools in a theorist's chest. They generate exact results when other methods fail and provide unexpected insights. The Kramers-Wannier duality \cite{Kramers1941} connects partition functions of the low and high-temperatures states of the Ising ferromagnet in $d=2$,  allowing an exact determination of the critical temperature. The mapping between the XY (easy-plane) ferromagnet and electrostatics in $d=2$ \cite{Kosterlitz1974} provided intuition about the interactions of magnetic vortices and underpinned the theory of the Kosterlitz-Thouless phase transition. Dualities between ferromagnets and gauge models in $d=3$ served as a window into the properties of gauge theories and helped understand the nature of quark confinement \cite{Kogut1979}.

In the analogy between the XY ferromagnet and electrostatics in $d=2$, vortices behave as electric charges. The definition of the vortex number $n$ as the increment of magnetization's azimuthal angle $\phi$ along the boundary of some region $\Omega$, $\int_{\partial \Omega} d \mathbf r \cdot \nabla \phi = 2\pi n$, can be recast as Gauss's law for the electric charge $Q$, $\int _{\partial \Omega} d \mathbf s \cdot \mathbf E = 2 \pi Q$, if we identify the vortex number with the electric charge,  $Q = n$, and the spatial gradients of the angle with components of an electric field, $E_i = \epsilon_{ij} \partial_j \phi$. Here Roman indices $i = 1,2$ refer to spatial directions and $\epsilon_{ij}$ is the Levi-Civita symbol in $d=2$. 

This duality has been extended to dynamical and quantum phenomena, which take place in a spacetime with $d=2+1$. The addition of the time dimension promotes electrostatics to electrodynamics, vortices become quantum particles with Bose statistics, and the XY ferromagnet is mapped to a superconductor interacting with an electromagnetic field \cite{Peskin1978, Fisher1989}. 

\begin{figure}
\includegraphics[width=0.99\columnwidth]{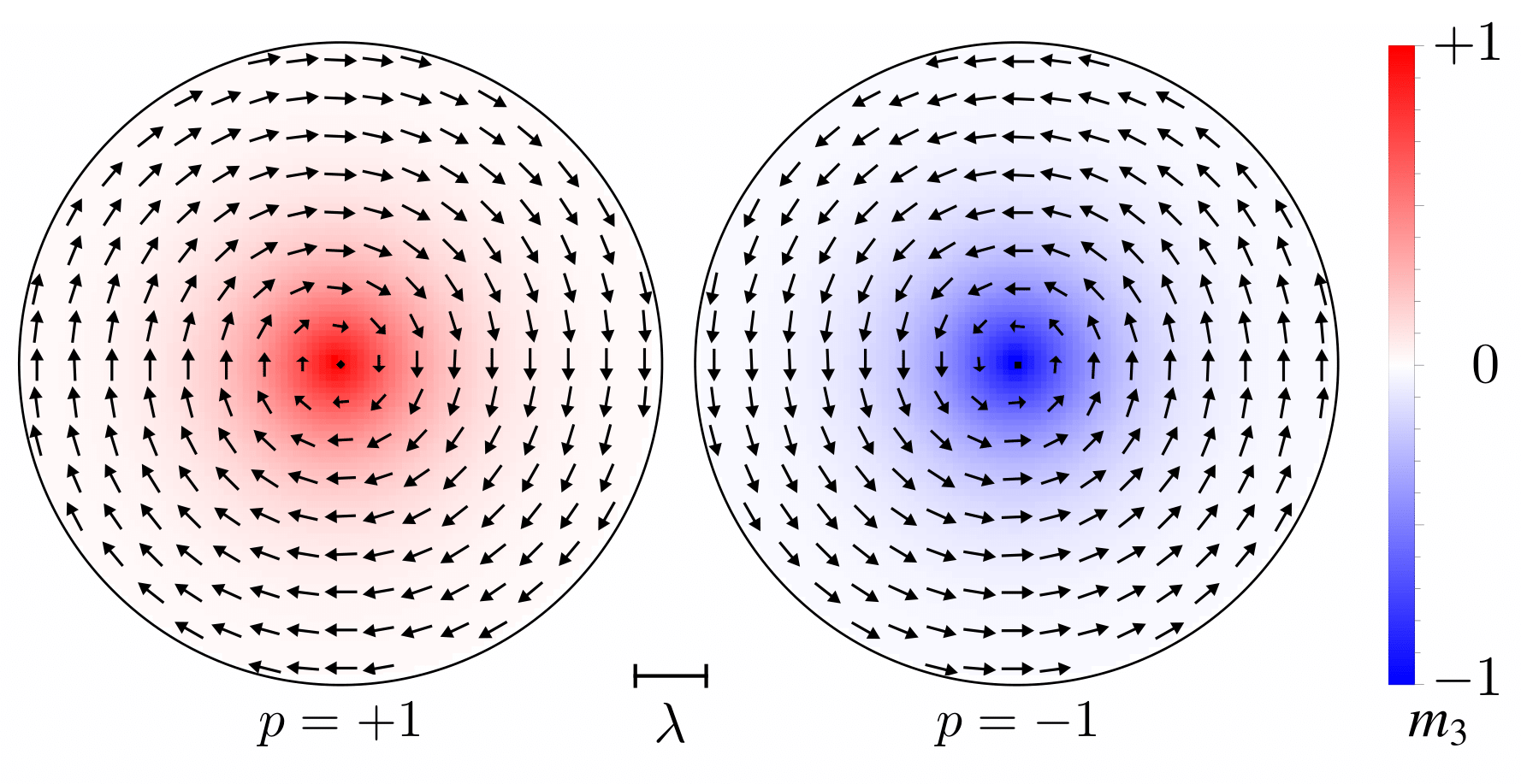}
\caption{Vortices in a thin film of permalloy. Numerical simulation in OOMMF \cite{oommf}. Color encodes $m_3$: positive (red), zero (white), and negative (blue). At a vortex core, magnetization leaves the easy plane and approaches the hard axis, $\mathbf m \to (0,0,p)$, where $p = \pm 1$ defines the polarity of the vortex.}
\label{fig:vortices}
\end{figure}

In this Letter, we revisit the duality between the XY ferromagnet and electrodynamics in $d=2+1$. In a realistic ferromagnet, the XY model with just two spin components represents a low-energy, long-wavelength limit of the Heisenberg ferromagnet with an easy-plane anisotropy. Although magnetization lies in the easy plane almost everywhere, it turns toward the hard axis at vortex cores, Fig.~\ref{fig:vortices}. Despite its small radius (typically a few nanometers \cite{Shinjo00, Wachowiak02}), the core plays a major role in the dynamics of a vortex. In particular, it is responsible for the gyroscopic (Magnus) force acting on a moving vortex \cite{Thiele1973, Huber1982, Choe2004}. This is a rare example where high-energy physics (here the existence of a vortex core) crucially impacts low-energy dynamics. 

The newly derived duality establishes an interesting connection between quantum statistics of vortices and the spin of the vortex core $S_3$ along the hard axis. In the dual description, vortices acquire not only the electric charge $Q = n$ but also a magnetic flux $\Phi = S_3$. \textcite{Wilczek1982a, Wilczek1982b} showed that in $d=2+1$ the quantum statistics of particles carrying both an electric charge $Q$ and a magnetic flux $\Phi$ is altered by the Aharonov-Bohm phase. Generally, bosons turn into anyons with the braiding phase $\vartheta = 2 \pi Q\Phi$. For magnetic vortices, this yields

\begin{equation}
\vartheta = 2 \pi n S_3.
\label{eq:braiding-phase-vortex}
\end{equation}
Simple vortices with $n = \pm 1$ and half-integer spin $S_3$ are therefore fermions. An even more exotic, anyon statistics is expected for vortices with a non-integer $2S_3$. 

Micromagnetics, the continuum theory of the easy-plane ferromagnet, operates with a unit-vector magnetization field 
\begin{equation}
\mathbf m 
    = (m_1, m_2, m_3)
    = (\sin{\theta} \cos{\phi},
        \sin{\theta} \sin{\phi},
        \cos{\theta}).
\label{eq:m}
\end{equation}
The simplest model without long-range dipolar interactions has the Lagrangian density
\begin{equation}
\mathcal L(\theta, \phi) = 
    \mathcal S (\cos{\theta} - p) \partial_t\phi
    - \mathcal{U}(\theta,\phi).
\label{eq:L-high-energy}
\end{equation}
The first term in the Lagrangian comes from the spin Berry phase and is responsible for the precessional dynamics of magnetization; $\mathcal S $ is the spin density. The number $p = \pm 1$ reflects a gauge choice and determines the location of a singularity of the spin wavefunction at $\cos{\theta} = - p = \mp 1$ \cite{Altland2010}. Either choice of $p$ would work if the spins stayed in the easy plane. However, a vortex configuration inevitably has a location where the spin orientation approaches one the poles, Fig.~\ref{fig:vortices}. To avoid the singularity, we have to make a specific choice of parameter $p$ \cite{Nikiforov1983} by equating it to the vortex polarity, defined as the value of the out-of-plane magnetization at the center of the vortex core, $m_3 = \pm 1$ \cite{Hellman2017}.

Neglecting long-range effects of the dipolar interaction, the potential energy has the area density  
\begin{equation}
\mathcal U(\theta,\phi) = 
    \frac{\mathcal A}{2} 
        [(\nabla \theta)^2 
        + \sin^2{\theta} (\nabla \phi)^2]
    + \frac{\mathcal K}{2} \cos^2{\theta}.
\label{eq:U-high-energy}
\end{equation}
Here $\mathcal A$ is the strength of Heisenberg exchange and $\mathcal K$ is the easy-plane anisotropy. The natural unit of length $\lambda = \sqrt{\mathcal A/\mathcal K}$ sets the size of a vortex core; the natural unit of time is $\tau =|\mathcal S|/\mathcal K$. The Lagrangian (\ref{eq:L-high-energy}) with the energy density (\ref{eq:U-high-energy}) represents a full (high-energy) theory of magnetization dynamics. 

In low-energy states, the magnetization field lies in the easy plane. The out-of-plane magnetization $m_3 = \cos{\theta} \ll 1$ is suppressed and can be viewed as a hard mode. In the spirit of the gradient expansion, we may neglect the $(\nabla\theta)^2$ term. With this simplification, the Lagrangian contains no gradients of the field $\theta$ and its (classical) equation of motion reads 
\begin{equation}
\mathcal K \cos{\theta} = \mathcal S \partial_t\phi.
\label{eq:eom-low-energy}
\end{equation}
In static equilibrium, $\partial_t\phi = 0$ and thus $\cos{\theta} = 0$, the magnetization resides strictly in the easy plane. Slow dynamics of the azimuthal angle $\phi$ is accompanied by a small tilt of magnetization out of the easy plane. The polar angle is thus a slave of the azimuthal angle. Integrating out $\theta$ from the action yields a low-energy theory with just one field $\phi$ and an effective Lagrangian
\begin{equation}
\mathcal L(\phi) = 
    -p \mathcal S \partial_t\phi
    + \frac{\rho}{2} (\partial_t\phi)^2 
    - \frac{\mathcal A}{2} (\nabla \phi)^2,
\label{eq:L-low-energy}
\end{equation}
where $\rho = \mathcal S^2/\mathcal K$ quantifies the inertia of the azimuthal angle. 

It is convenient to write the Lagrangian in a Lorentz-covariant form with the Minkowski metric $\eta_{\mu\nu} = \text{diag}(+1,-1,-1)$ and in natural units, 
\begin{equation}
\mathcal L(\phi) = 
    \bar{\sigma}^\mu \partial_\mu \phi
    + \frac{e^2}{2}\partial_\mu \phi \,
        \partial^\mu \phi,
\quad
\bar{\sigma}^\mu = - p e^2 \delta^\mu_0.
\label{eq:L-low-energy-natural-units}
\end{equation}
The dimensionless coupling constant $e^2 \equiv |\mathcal S| \mathcal A /\mathcal K \gg 1$ is roughly the net out-of-plane spin $S_3$ of a vortex core. 

The low-energy Lagrangian (\ref{eq:L-low-energy}) has a global symmetry of rotations in the easy plane, $\phi \mapsto \phi + \text{const}$. The conserved global quantity is the hard-axis spin component $S_3$. The associated local conservation law, $\partial_\mu \sigma^\mu = 0$, is the continuity equation for the spin current defined as
\begin{equation}
\sigma^\mu
    \equiv \frac{\partial \mathcal L}
    {\partial (\partial_\mu \phi)}  
    - \bar{\sigma}^\mu = 
        e^2 \partial^\mu \phi.
\label{eq:spin-current-low-energy}
\end{equation}
Here we have separated a uniform background spin current $\bar{\sigma}$, whose only nonvanishing component $\bar{\sigma}^0 = - p \mathcal S$ is a background spin density, from the dynamical part $\sigma$. Although the linear term $\bar{\sigma}^\mu \partial_\mu \phi$ in the Lagrangian (\ref{eq:L-low-energy-natural-units}) does not influence the classical equation of motion,
\begin{equation}
\partial_\mu \partial^\mu \phi = 0,   \label{eq:wave-equation-low-energy}
\end{equation}
it has a topological character and plays an important role in the dynamics of vortices, as we discuss below. Eq.~(\ref{eq:wave-equation-low-energy}) describes spin waves with a linear dispersion, $\omega = k$. \\

Next we derive the dual theory of electrodynamics by starting with the effective low-energy model (\ref{eq:L-low-energy-natural-units}). Although this duality is well known in field theory (see \textcite{Tong} for a pedagogical review), we will use the occasion to illustrate the underlying ideas that will be useful for obtaining the dual description of the full model (\ref{eq:L-high-energy}).

The duality can be revealed most efficiently in the language of differential forms. In $d=2+1$ the electromagnetic field is represented by a 2-form $F = \frac{1}{2}F_{\mu\nu} dx^\mu \wedge dx^\nu$ and the electric current by a 1-form $J = J_\mu dx^\mu$ \cite{Misner1973}. Maxwell's equations and current conservation read
\begin{subequations}
\begin{equation}
d \hodge{F} = 2\pi \hodge{J}, 
\quad
dF = 0, 
\quad
d\hodge{J} = 0.
\label{eq:EM-diff-forms}
\end{equation}
Here $d$ is the exterior derivative and $\hodge$ is the Hodge dual. In the theory of the XY ferromagnet, the spin and vortex currents are represented by 1-forms $\sigma$ and $j$. The relation between them, and the conservation of the two currents read (in the low-energy limit)
\begin{equation}
d\sigma = 2\pi e^2 \hodge{j},
\quad
d\hodge{\sigma} = 0,
\quad
d\hodge{j} = 0.
\label{eq:XY-diff-forms}
\end{equation}
\label{eq:eom-diff-forms}
\end{subequations}
Comparing Eqs.~(\ref{eq:eom-diff-forms}) shows that the vortex current $j$ maps to the electric current $J$ and the spin current $\sigma$ to the Hodge dual of the electromagnetic field $\hodge{F}$.

We unpack this analogy in the more familiar language of tensors and components, beginning with a list of ingredients expected in a theory of electrodynamics:

\emph{Gauge field.} An electromagnetic field should satisfy local constraints (Bianchi identities) in the form of the homogeneous Maxwell equations. These constraints are resolved by expressing the electromagnetic field as the curl of a gauge field, $F_{\mu\nu} = \partial_\mu A_\nu - \partial_\nu A_\mu$. The Bianchi identity in $d=2+1$ reads 
\begin{equation}
\partial_\mu \hodge{F}^\mu = 0,
\quad
\hodge{F}^\mu \equiv \frac{1}{2} \epsilon^{\mu\nu\rho} F_{\nu\rho},
\label{eq:Bianchi-identity}
\end{equation}
where $\epsilon^{\mu\nu\rho}$ is the Levi-Civita symbol in $d=2+1$. Here $\hodge{F}$ is the Hodge dual of the electromagnetic field $F$ \cite{Misner1973, Jackson1975}.  It corresponds to a conserved current for a global $U(1)_J$ symmetry, referred to as topological $U(1)$, which exist for Maxwell theories in $d=2+1$.  The theory admits monopole defect operators charged under $U(1)_J$. 

The global symmetry in the ferromagnetic model is the symmetry of spin rotations in the $xy$ plane. We identify the generator of this symmetry with that of the $U(1)_J$ of the Maxwell theory, and thus the current $\sigma^\mu$ maps to $\hodge{F}^\mu$ as follows:
\begin{equation}
\hodge{F}^\mu \equiv - \sigma^\mu 
= -e^2 \partial^\mu \phi,
\quad
\hodge{\bar{F}}^\mu \equiv - \bar{\sigma}^\mu.
\label{eq:duality-low-energy}
\end{equation}
Here quantities with a bar represent uniform background parts of the respective fields. The minus signs in Eq.~(\ref{eq:duality-low-energy}) reflect the convention that a positive vortex number corresponds to a positive electric charge.  

With the physical units restored, the electric and magnetic fields are
\begin{equation}
E^i = \mathcal A \epsilon^{ij} \partial_j \phi,
\quad
B = \rho \partial_t\phi,
\quad 
\bar{B} =  -p \mathcal S.
\label{eq:EM-field-low-energy}
\end{equation}
As in $d=2$ \cite{Kosterlitz1974}, the electric field comes from spatial gradients of $\phi$, whereas the temporal gradient gives rise to the magnetic field. The background magnetic field $\bar{B} = - p \mathcal S$ represents an effect well known in vortex dynamics. A particle with electric charge $Q$ moving with velocity $\dot{x}^i$ should experience the Lorentz force $F_i = 2\pi Q \bar{B} \epsilon_{ij} \dot{x}^j$ \cite{twopi}. With $Q = n$ and $\bar{B} = - p \mathcal S$, this exactly reproduces the gyroscopic force $F_i = - 2\pi n p \mathcal S \epsilon_{ij} \dot{x}^j$ acting on a moving vortex \cite{Thiele1973, Huber1982, Nikiforov1983}. 

\emph{Electromagnetic waves.} A hallmark of Maxwell's theory is the existence of transverse electromagnetic waves with a linear dispersion, $\omega = k$. Spin waves in the XY ferromagnet (\ref{eq:wave-equation-low-energy}) seem like a good candidate. There is just one spin-wave mode for each wavevector, in accordance with a single transverse polarization expected for electromagnetic waves in $d=2+1$. The transverse nature of the electric field in a spin wave can be checked with the aid of  Eq.~(\ref{eq:EM-field-low-energy}):
$\partial_i E^i
    = \mathcal A(\partial_x \partial_y  
        - \partial_y \partial_x) \phi
    = 0$
in the absence of vortices. 

\emph{Coupling of the field and current.} To find a conserved matter current satisfying the continuity equation, $\partial_\mu j^\mu = 0$, we turn to vortices. They are indestructible and can only be annihilated in pairs.
In their presence, derivatives of $\phi$ are singular, $\partial_x \partial_y \phi - \partial_y \partial_x \phi = 2\pi \rho$. This definition of vortex density $\rho$ generalizes to a vortex current $j^\mu$ in $d=2+1$:

\begin{equation}
\epsilon^{\mu\nu\rho} \partial_\nu \partial_\rho \phi = 2\pi j^\mu.    
\end{equation}
With the help of the duality relation (\ref{eq:duality-low-energy}), this identity takes the form of the inhomogeneous Maxwell equations,
\begin{equation}
\partial_\mu F^{\mu\nu} = 2 \pi e^2 J^{\nu},
\label{eq:Maxwell-equations-low-energy}
\end{equation}
with the electric current $J$ equal to the vortex current $j$. The dual theory can be obtained from the Lagrangian of Maxwell's electrodynamics with a matter current $J$ coupled to both the dynamical and background gauge fields $A$ and $\bar{A}$:

\begin{equation}
\mathcal L(J,A) =  
-2\pi (A_\mu + \bar{A}_\mu)J^\mu
- \frac{F_{\mu\nu} F^{\mu\nu}}{4 e^2}.
\label{eq:L-EM-low-energy-natural-units}
\end{equation}

\emph{Duality via an auxiliary field.} We now derive the dual theory (\ref{eq:L-EM-low-energy-natural-units}) from the low-energy Lagrangian (\ref{eq:L-low-energy-natural-units}) in a standard formal way \cite{Tong}, through the introduction of an auxiliary vector field with components $\hodge{F}^\mu$. The Lagrangian of the two fields $\phi$ and $\hodge{F}$ is chosen to be 
\begin{equation}
\mathcal L(\phi, \hodge{F}) = 
-(\hodge{F}^\mu + \hodge{\bar{F}}^\mu) 
    \partial_\mu \phi
- \frac{\hodge{F}^\mu \hodge{F}_\mu}{2 e^2}.   
\label{eq:L-varphi-a-F}
\end{equation}
This choice assures that minimization of the action with respect to $\hodge{F}$ yields the conjectured relation (\ref{eq:duality-low-energy}). Integrating out the auxiliary field $\hodge{F}$ would lead to our effective theory (\ref{eq:L-low-energy-natural-units}). Instead, we will keep the auxiliary field $\hodge{F}$ and integrate out the angle field $\phi$. 

However, prior to that, we need to separate a singular vortex part of the field $\phi$ from spin waves as it is done in $d=2$ \cite{Kosterlitz1974}. In the presence of vortices, $\phi$ is not a single-valued function of the spacetime coordinates and $\partial_\mu \phi$ is not, strictly speaking, a gradient. We separate this quantity into two parts, $\partial_\mu \phi = a_\mu + \partial_\mu \varphi$. The new gauge field $a$ is defined by vortex world-lines, 

\begin{equation}
\epsilon^{\mu\nu\rho} \partial_\nu a_\rho  
    = 2\pi j^\mu.
\label{eq:vortex-current}
\end{equation}
The single-valued field $\varphi$ represents spin waves in the original theory and generates gauge transformations for the vortex gauge field $a$. 

Integrating out the single-valued part of the field $\varphi$ produces the Bianchi identity for $F$ (\ref{eq:Bianchi-identity}). Upon resolving it in the expected way, $\hodge{F}^\mu = \epsilon^{\mu\nu\rho} \partial_\nu A_\rho$, we obtain the Lagrangian for a gauge field $A$ and the vortex current $j$ parametrized by the vortex gauge field $a$: 
\begin{equation}
\mathcal L(j, A) = 
-\epsilon^{\mu\nu\rho} 
    a_\mu \partial_\nu (A_\rho + \bar{A}_\rho)
- \frac{F_{\mu\nu} F^{\mu\nu}}{4e^2}.
\label{eq:L-a-A}
\end{equation}
Note that the first term in Eq.~(\ref{eq:L-a-A}) is $a_\mu \sigma^\mu$, indicating that the role of the electric charge for the gauge field $a$ is played by the spin $S_3$, whereas the electric charge for $A$ is the vortex number $n$. 

Finally, we convert the first term in Eq.~(\ref{eq:L-a-A}) via integration by parts and use the relation between $a$ and $j$ (\ref{eq:vortex-current}) to obtain the conjectured Lagrangian of the dual theory (\ref{eq:L-EM-low-energy-natural-units}). 

As already mentioned, the low-energy theory (\ref{eq:L-low-energy-natural-units}) and its well-known dual (\ref{eq:L-EM-low-energy-natural-units}) break down at vortex cores. We now turn to the full model (\ref{eq:L-high-energy}) and derive its hitherto unknown dual theory (\ref{eq:L-EM-high-energy-natural-units}), our main technical achievement underpinning the new conceptual result (\ref{eq:braiding-phase-vortex}). 

We can readily construct the electromagnetic fields following the familiar route. The Lagrangian (\ref{eq:L-high-energy}) and  potential energy (\ref{eq:U-high-energy}) retain the global rotational symmetry. The spin current $\sigma^\mu$ has the following components: 

\begin{equation}
\sigma^0 = \mathcal S \cos{\theta},
\quad
\bar{\sigma}^0 = - p \mathcal S,
\quad 
\sigma^i = - \mathcal A \sin^2{\theta} \, \partial_i \phi.
\end{equation}
The dynamical temporal component $\sigma^0$ is the density of spin along the hard axis. Identification of the spin current with the electromagnetic field along the lines of Eqs.~(\ref{eq:duality-low-energy}) and (\ref{eq:EM-field-low-energy}) yields 
\begin{equation}
E^i = \mathcal A \sin^2{\theta} \, \epsilon^{ij} \partial_j \phi,
\quad
B = \mathcal S \cos{\theta},
\quad
\bar{B} = - p\mathcal S.
\label{eq:EM-field-high-energy}
\end{equation}
The low-energy result (\ref{eq:EM-field-low-energy}) is recovered if we set $\sin{\theta} = 1$ and use the low-energy equation of motion (\ref{eq:eom-low-energy}).

For completeness, we give the Lagrangian of the dual gauge theory, in natural units:
\begin{equation}
\mathcal L(J,A) =
-2\pi (A_\mu + \bar{A}_\mu) J^\mu 
+ \frac{1}{2 e^2} \left(
    \frac{\mathbf E \cdot \mathbf E - (\nabla B)^2}{1 - (B/\bar{B})^2}
    - B^2
\right).
\label{eq:L-EM-high-energy-natural-units}
\end{equation}
The Lorentz-covariant form (\ref{eq:L-EM-low-energy-natural-units}) is recovered in the limit when the dynamical magnetic field is weak and varies slowly in space, $\nabla B \ll B \ll \bar{B}$. 

Up to this point, our theory of the XY ferromagnet in $d=2+1$, recast as electrodynamics, has faithfully reproduced what is already known. The electrostatic analogy goes back to 1974 \cite{Kosterlitz1974}; the dynamical similarity with electric charges in a background magnetic field is also not new \cite{Papanicolaou1991, Baryakhtar1994, Ivanov1998}. Does this duality provide any new insights, not obvious from the original theory? 

One interesting feature that, as far as we know, has not been previously pointed out is the presence of a magnetic field $B = \mathcal S \cos{\theta}$ localized at a vortex core, where $\cos{\theta} \neq 0$. The net magnetic flux of a vortex,
\begin{equation}
\Phi = \int d^2x \, B 
    = \int d^2x \, \mathcal S \cos{\theta}
    = S_3,
\end{equation}
is equal to the net spin $S_3$ of the vortex core. We thus find that a vortex behaves like a particle with both an electric charge $Q = n$ and a magnetic flux $\Phi = S_3$. The attachment of a well-localized magnetic flux does not influence the classical dynamics of a charged particle. However, it has important consequences at the quantum level because of the Aharonov-Bohm phase experienced by an electric charge moving around a magnetic flux. \textcite{Wilczek1982a, Wilczek1982b} pointed out that particles carrying both an electric charge $q$ and magnetic flux $\Phi$ in $d=2+1$, increment their statistical angle $\vartheta$ (0 for bosons and $\pi$ for fermions) by $2\pi Q \Phi$ (in our units \cite{twopi}). Viewed as a quantum particle, a vortex in a ferromagnet is ordinarily considered to be a boson \cite{Fisher1989}. The idea that a vortex carries both an electric charge $Q = n$ and a magnetic flux $\Phi = S_3$ means that its statistical angle is $\vartheta = 2\pi n S_3$. Common single vortices ($n = \pm 1$) can exhibit the fermion statistics if their spin $S_3$ is half-integer. 

Are there vortices with a half-integer spin $S_3$? We do not know for sure. It is relatively easy to determine the spin of a vortex in a classical model such as the one defined by Eq.~(\ref{eq:U-high-energy}). The vortex core is well defined and its net spin is of the order of $e^2 = |\mathcal S| \mathcal A/\mathcal K \gg 1$. However, this classical answer varies continuously with the parameters of the model and is not quantized. 

The problem needs to be solved at the quantum level. Aside from technical difficulties, we encounter a conceptual problem. The transverse spin $S_3$ is a conserved quantity by virtue of the $O(2)$ rotational symmetry. However, in an ordered ferromagnet this symmetry is spontaneously broken. Therefore, the ground state of an ordered magnet is generally a superposition of (infinitely) many states with different values of $S_3$, 

\begin{equation}
|\psi \rangle = \sum_{S_3} C_{S_3} |S_3\rangle,
\label{eq:psi-superposition}
\end{equation}

and $S_3$ is not even a well-defined quantity. Fortunately, quantum statistics is determined not so much by the statistical angle $\vartheta$ but by its exponential $e^{i \vartheta} = e^{2\pi i n S_3}$.  Because physical states are invariant under $2\pi$ rotations, the superposition (\ref{eq:psi-superposition}) may only contain values of $S_3$ differing by integers, e.g., 1/2, 3/2, 5/2, \ldots or 0, 1, 2, \ldots. The number $e^{2\pi i n S_3}$ is the same for all such $S_3$, so the quantum statistics of vortices is well defined even if $S_3$ is not. 

We speculate that vortices with a half-integer spin could be found in single-layer ferromagnets. With two layers, the total spin would presumably double and give the trivial bosonic statistics. For the same reason, magnetic atoms with half-integer spin look more promising than ones with integer spin. 

The attachment of fluxes to charges is absent in the naive dual theory (\ref{eq:L-EM-low-energy-natural-units}). One could attempt to fix this deficiency by adding a Chern-Simons (CS) term, $\mathcal L_\text{CS} = \pi k \, \epsilon^{\mu\nu\rho} A_\mu \partial_\nu A_\rho$ \cite{Referee-C,Dasgupta2018}. Doing so would not affect the classical dynamics \cite{Tong} and attach magnetic flux $\Phi = Q/k$ to an electric charge $Q$. However, this one-to-one correspondence between the charge and flux is too restrictive for our model. A magnetic vortex with ``electric charge'' $Q = n$ can have both positive and negative transverse ``magnetic flux'' $\Phi = S_3$, depending on the polarity $p = \pm 1$ of the core. This $\mathbb{Z}_2$ degree of freedom is missing in the standard scenario of flux attachment via a CS term, thus requiring a more sophisticated approach.

Vortices in ferromagnets have been extensively studied, both experimentally and theoretically. In practically all of these studies, vortices have been treated as \emph{classical} objects. Only recently have theorists begun to ponder their unusual \emph{quantum} properties. For example, Ivanov and co-workers \cite{Galkin2007, Ivanov2010} considered the quantum mechanics of a single vortex in an atomic lattice with spins of length $S$. The single-vortex energy spectrum consists of $2S$ bands reminiscent of electron bands in a solid. Similar results  for skyrmion energy bands were obtained by \textcite{Takashima2016}. Noncommutativity of momentum components for vortices and skyrmions was pointed out by \textcite{Watanabe2014}; the same applies to their coordinates \cite{Tchernyshyov2015}.

In this Letter, we have shown that magnetic vortices, viewed as quantum particles, may exhibit nontrivial quantum statistics: vortices with a half-integer core spin $S_3$ are expected to be fermions. Even more exotic anyon statistics is expected for vortices with a non-integer $2S_3$. The existence of vortices with non-integer $2S_3$, also conjectured independently by \textcite{Ivanov-private}, would be a tantalizing possibility. However, \textcite{Feldman-private} has pointed out that anyon statistics can probably be ruled out for vortices on account of the spin--statistics theorem \cite{Wilczek1982a, Preskill}, which sets $e^{i\vartheta} = e^{-2\pi i S_3}$. This result is compatible with Eq.~(\ref{eq:braiding-phase-vortex}) for $n=1$ only if $2S_3$ is an integer.

We hope that our work will stimulate further interest in quantum mechanics of vortices and other magnetic solitons. 

\begin{acknowledgments}
\emph{Acknowledgments.} We are grateful to Dmitri Feldman, Martin Greiter, and Boris Ivanov for illuminating discussions. S. D., S. Z., and O. T. have been supported by the U.S. DOE Basic Energy Sciences,Materials Sciences and Engineering Award DE-SC0019331. I. B. has been supported by the NSF Grant PHY-1820784. O. T. acknowledges the hospitality of the Aspen Center for Physics, which is supported by National Science Foundation Grant PHY-1607611 and of the Kavli Institute for Theoretical Physics. S. Z. acknowledges the Graduate Fellowship Program at the Kavli Institute for Theoretical Physics, which is funded by the National Science Foundation under Grant PHY-1748958 and by the Heising-Simons Foundation.
\end{acknowledgments}

\bibliographystyle{apsrev4-1}
\bibliography{main}

\begin{thebibliography}{33}
\bibitem [{Ref()}]{Referee-C}%
  \BibitemOpen
  \href@noop {} {}\bibinfo {note} {We owe this idea to one of the referees of
  this paper}\BibitemShut {NoStop}%
\end{thebibliography}

\clearpage
\onecolumngrid

\begin{center}
  \large{\textbf{Supplemental Material}}  
\end{center}
\appendix
\setcounter{equation}{0}
\renewcommand{\theequation}{A.\arabic{equation}}

\section{Electromagnetism in 2+1 dimensions}

We give a brief summary of electromagnetism in a Minkowski spacetime with $d=2+1$ dimensions along the lines of \textcite{Jackson1975}. The metric tensor is $\eta_{\mu\nu} = \text{diag}{(+1,-1,-1)}$. 

The electromagnetic field has three components, the magnetic field $B$ and the electric field $\mathbf E = (E_x, E_y)$. The gauge field also has three components, the electrostatic potential  $\phi$ and the vector potential $\mathbf{A} = (A_x, A_y)$. 

In the relativistic notation, the gauge field has the following covariant and contravariant components: 
\begin{equation}
A_\mu =     \left(
\begin{array}{c}
\phi\\
-A_x\\
-A_y
\end{array}
\right),
\quad
A^\mu = \left(
\begin{array}{c}
\phi\\
A_x\\
A_y
\end{array}
\right).
\end{equation}
The electromagnetic field is an antisymmetric tensor $F_{\mu\nu} = \partial_{\mu}A_{\nu} - \partial_{\nu}A_{\mu}$:
\begin{equation}
F_{\mu\nu} = \left(
\begin{array}{ccc}
 0 & E_x & E_y \\
 -E_x & 0 & -B \\
 -E_y & B & 0
\end{array}
\right), 
\quad
F^{\mu\nu} = \left(
\begin{array}{ccc}
 0 & -E_x & -E_y \\
 E_x & 0 & -B \\
 E_y & B & 0
\end{array}
\right).
\label{eq:Electromagnetic-tensor-2+1}
\end{equation}
The field strength tensor $F_{\mu\nu}$ can also be represented by its dual \cite{Misner1973, Jackson1975}, which in $d=2+1$ is a 3-vector
\begin{equation}\label{eq.Hodge_dual}
 \hodge{F}^{\mu} = \frac{1}{2}\epsilon^{\mu\nu\rho}F_{\nu\rho} 
 = 
 \left(
 \begin{array}{c}
 -B\\
 -E_y\\
 E_x
 \end{array}
 \right).   
\end{equation}

The homogeneous Maxwell equation, $\partial_x E_y - \partial_y E_x + \partial_t B = 0$, reads $\partial_{\mu}\hodge{F}^{\mu} = 0$ in the relativistic notation and is resolved by expressing the dual field as a 3-curl of the gauge field, $\hodge{F}^\mu = \epsilon^{\mu\nu\rho} \partial_\nu A_\rho$, or $F_{\mu\nu} = \partial_\mu A_\nu - \partial_\nu A_\mu$. 

The inhomogenous Maxwell equations, $ \partial_{\mu}F^{\mu\nu} = 2\pi j^{\nu} $, can be derived from the Lagrangian: 
\begin{equation}
    \mathcal{L}(A,j) = - A_{\mu}  j^{\mu} -\frac{1}{8\pi}F^{\mu\nu}F_{\mu\nu}.
\label{eq:L-EM-2+1}
\end{equation}
The second term in the Lagrangian (\ref{eq:L-EM-2+1}) represents the kinetic and potential energy densities of the electromagnetic field, $\mathbf E \cdot \mathbf E/(4\pi)$ and $B^2/(4\pi)$, respectively. The first term expresses the coupling between the electromagnetic field and electric current. For a point particle with spacetime coordinates $x^\mu$ and electric charge $q$, it generates the action term 
\begin{equation}
S  = - q \int A_\mu dx^\mu 
    = q \int 
        (\mathbf A \cdot d \mathbf r 
        - \phi \, dt).
\end{equation}
This action term is responsible for the 3-force $f_\mu = q F_{\mu\nu}\dot{x}^\nu$, where the dot means the derivative with respect to proper time $\tau$, $d\tau^2 = dx^\mu dx_\mu$. Its spatial components, $f_x = q B \dot{y}$ and $f_y = - q B \dot{x}$, represent the Lorentz force.

\section{Obtaining the duality through differential forms}
\noindent
We present a derivation of the dual theory contained in the Lagrangian density, Eq.(\ref{eq:L-a-A}) in the language of differential geometry where the Hodge dual of the electromagnetic field, Eq.(\ref{eq.Hodge_dual}) is explicitly manifest.\\
To start the construction we note that the two conserved currents introduced in the text $(\sigma^{\mu},j^{\mu})$, representing the conservation of spin and vortex number, can be written into 2-forms $(\hodge{\sigma},\hodge{j})$ such that the conservation laws are expressed as continuity equations $d(\hodge{\sigma}) = 0$ and $d(\hodge{j}) = 0$.\\
This provides a natural setting to introduce the 2-form EM tensors $F$ and $f$ through the Bianchi identities, $dF = 0$ and $df = 0$. Starting from Eq.(\ref{eq:L-varphi-a-F}), written in terms of the two form $F$:
\begin{equation}
    \mathcal{L} = -(F + \bar{F})\wedge(d\varphi + a) - \frac{F\wedge\hodge{F}}{4 e^2}
\end{equation}
where $F$ and its dual $\hodge{F}$ are still auxiliary fields and integrating them out reproduces the theory of spin waves. Here $\varphi$ is the slowly varying non singular spin wave field, and the singular field $a$ is mapped to the vortex current:
\begin{eqnarray}\label{eq.vortex-current}
    da &=& 2\pi e^2\hodge{j} \\ \nonumber
    d(\hodge{j}) &=& \frac{1}{2\pi e^2}d^{2}a = 0.
\end{eqnarray}
where $j$ is the conserved vortex number current as expressed through the continuity equation.\\
Now the conservation of spin current, $d\hodge{\sigma} = 0$, and its identification with the auxiliary 2-form field $-F$, see Eq.(\ref{eq:duality-low-energy}), imposes $d(F + \bar{F}) = 0$ (Bianchi identity). This can be expressed by introducing gauge fields $(A,\bar{A})$ such that $F + \bar{F} = d(A + \bar{A})$. In a similar vein the continuity equation for the vortex current $d\hodge{j} = 0$ provides the Bianchi identity for the 2-form $f$. Integrating out the slowly varying, non singular $\varphi$ field we obtain the Lagrangian density:
\begin{equation}
    \mathcal{L} = - a\wedge d(A + \bar{A}) - \frac{F\wedge \hodge{F}}{4e^2} = -2\pi(A + \bar{A})\wedge(\hodge{j}) - \frac{F\wedge \hodge{F}}{4 e^2},
\end{equation}
where we used the vortex current equation, Eq.(\ref{eq.vortex-current}) and an integration by parts to obtain the final form. This is the same theory as Eq.(\ref{eq:L-a-A}) expressed using differential forms. For the high energy version a similar construction where the conserved current 2-forms are mapped to the EM tensor 2-forms goes through.

\section{Dual theory of the XY ferromagnet}
In the dual theory, we have identified the spin current $\sigma^\mu$ with the electromagnetic field $\hodge{F}^\mu$ in Eq.~(\ref{eq:duality-low-energy}). The minus sign in $\sigma^\mu = - \hodge{F}^\mu$ ensures the same sign for the vortex number, 
\begin{equation}
n = \frac{1}{2\pi} \int_{\partial \Omega} d \mathbf r \cdot \nabla \phi,    
\end{equation}
and its dual electric charge, 
\begin{equation}
Q = \frac{1}{2\pi} \int_{\partial \Omega} d \mathbf s \cdot \mathbf E.    
\end{equation}
Indeed, by using the expressions for the electric field, Eqs.~(\ref{eq:EM-field-low-energy}) or (\ref{eq:EM-field-high-energy}), we obtain $Q = \mathcal An$, where $\mathcal A$ is the Heisenberg exchange constant in Eq.~(\ref{eq:U-high-energy}).

The choice of the proportionality coefficient between the spin current and the dual electromagnetic field in Eq.~(\ref{eq:duality-low-energy}) fixes the strength of the coupling between the electric current and the gauge field in the dual theory, Eqs.~(\ref{eq:L-EM-low-energy-natural-units}) and (\ref{eq:L-EM-high-energy-natural-units}). The coupling term $\mathcal L = - 2\pi A_\mu j^\mu$ bears an extra factor of $2\pi$ compared to regular electrodynamics in $d=2+1$ (\ref{eq:L-EM-2+1}). For this reason, both the Aharonov-Bohm phase and the Lorentz force have an extra factor of $2\pi$. 

\end{document}